\documentclass[10pt,conference]{IEEEtran}
\IEEEoverridecommandlockouts
\usepackage{amsmath,amssymb,amsfonts}
\usepackage{graphicx}
\usepackage{textcomp}
\def\BibTeX{{\rm B\kern-.05em{\sc i\kern-.025em b}\kern-.08em
    T\kern-.1667em\lower.7ex\hbox{E}\kern-.125emX}}

\usepackage{tabularx}
\usepackage{threeparttable}
\usepackage{colortbl}
\usepackage{booktabs}   

\makeatletter
\let\MYcaption\@makecaption
\makeatother

\usepackage[font=footnotesize]{subcaption}
\usepackage{zi4} 

\makeatletter
\let\@makecaption\MYcaption
\makeatother
        
\newif\iffull
\fullfalse      
                        
\usepackage{balance}

\usepackage[utf8]{inputenc} 
\usepackage[T1]{fontenc}
\usepackage{microtype}

\newcommand{\rone}{(\emph{i})~}
\newcommand{\rtwo}{(\emph{ii})~}

\usepackage{url}
\usepackage{enumitem}
\newcommand\twodigits[1]{%
  \ifnum#1<10 \number#1 \else #1\fi
}
\pdfoutput=1
\usepackage{color}
\usepackage{listings}
\usepackage{amssymb}
\usepackage{multirow}
\usepackage{etoolbox}
\usepackage[plain]{algorithm}
\usepackage[noend]{algorithmic}
\usepackage{syntax}
\usepackage{comment}
\usepackage{amsthm}
\usepackage{amsmath}
\usepackage{xspace}
\usepackage[svgnames]{xcolor}
\usepackage{tcolorbox}
\usepackage{tikz}
\usepackage[all]{nowidow}
\usepackage[english]{babel}
\usepackage{hyphenat}
\usepackage{listings}
\usepackage{lstlinebgrd}
\definecolor{pblue}{rgb}{0.13,0.13,1}
\definecolor{pgreen}{rgb}{0,0.5,0}
\definecolor{pred}{rgb}{0.9,0,0}
\definecolor{pgrey}{rgb}{0.46,0.45,0.48}
\definecolor{darkblue}{rgb}{0.0, 0.0, 0.55}
\makeatletter
\newcommand{\ALOOP}[1]{\ALC@it\algorithmicloop\ #1%
	\begin{ALC@loop}}
	\newcommand{\ENDALOOP}{\end{ALC@loop}\ALC@it\algorithmicendloop}

\usepackage{setspace}
\let\Algorithm\algorithm
\renewcommand\algorithm[1][]{\Algorithm[#1]\setstretch{1}}

\usepackage{newfloat}
\DeclareFloatingEnvironment[fileext=lop]{editpattern}

\usepackage[capitalize]{cleveref}

\usepackage{cite}
\usepackage{float}
\usepackage{makecell}
\usepackage{microtype}
\usepackage{amsmath}
\usepackage{mdframed}

\usepackage[strict]{changepage}

\usepackage{framed}
\makeatletter
\renewenvironment{framed}{%
 \def\FrameCommand##1{\hskip\@totalleftmargin
 \fboxsep=\FrameSep\fbox{##1}
     \hskip-\linewidth \hskip-\@totalleftmargin \hskip\columnwidth}%
 \MakeFramed {\advance\hsize-\width
   \@totalleftmargin\z@ \linewidth\hsize
   \@setminipage}}%
 {\par\unskip\endMakeFramed}
\makeatother

\definecolor{formalshade}{rgb}{0.95,0.95,1}

\newenvironment{formal}{%
  \def\FrameCommand{%
    \hspace{1pt}%
    {\color{darkblue}\vrule width 2pt}%
    {\color{formalshade}\vrule width 4pt}%
    \colorbox{formalshade}%
  }%
  \MakeFramed{\advance\hsize-\width\FrameRestore}%
  \noindent\hspace{-6pt}
  \begin{adjustwidth}{}{7pt}%
  \vspace{2pt}\vspace{2pt}%
}
{%
  \vspace{3pt}\end{adjustwidth}\endMakeFramed%
}

\newcounter{resultcounter}
\newenvironment{result}{\begin{formal}
	\refstepcounter{resultcounter}
}{
\end{formal}
}

\newcounter{patterncounter}

\newtheorem{definition}{Definition}
\newtheorem{theorem}{Theorem}

\usepackage[all]{nowidow}
\usepackage{etoolbox}
\patchcmd{\quote}{\rightmargin}{\leftmargin 1em \rightmargin}{}{}
                        
\newtoggle{release}
\newtoggle{full}

\togglefalse{release}
\toggletrue{full}

\newcommand{\reudismam}[1]{}

\newtoggle{highlightchanges}
\togglefalse{highlightchanges}

\iftoggle{highlightchanges}{
   \newcommand {\changes}[1]{{\color{purple}{#1}\normalfont}}
}{
  \newcommand {\changes}[1]{{#1}}
}

\newtoggle{highlightwrong}
\toggletrue{highlightwrong}

\iftoggle{highlightwrong}{
   
}{
  
}

\usepackage{expl3,xparse}
\ExplSyntaxOn
\NewDocumentCommand \lstcolorlines { O{green} m }
{
 \clist_if_in:nVT { #2 } { \the\value{lstnumber} }{ \color{#1} }
}
\ExplSyntaxOff

\usepackage{lstlinebgrd}

\definecolor{light-gray}{gray}{0.92}

\definecolor{diffstart}{named}{Grey}
\definecolor{diffincl}{named}{DarkBlue}
\definecolor{diffrem}{named}{Red}

\lstdefinelanguage{diff}{
    basicstyle=\ttfamily\footnotesize,
    morecomment=[f][\color{diffstart}]{@@},
    morecomment=[f][\color{diffincl}]{+\ },
    morecomment=[f][\color{diffrem}]{-\ },
    morekeywords={def, while, return, if, foreach}
}

\lstdefinelanguage{trans}{
  basicstyle=\footnotesize\tt,        
  breakatwhitespace=false,         
  breaklines=true,                 
  captionpos=b,                    
  extendedchars=true,              
  frame=single,                    
  language=Java,                 
  keywordstyle=\bf,
  showspaces=false,                
  showstringspaces=false,          
  showtabs=false,                  
  tabsize=2                       
}

\def\code#1{\texttt{#1}}

\newcommand{\noise}{295}
\newcommand{\duplicates}{109}
\newcommand{\technique}{\textsc{Revisar}}
\newcommand{\quickpl}{quick fixes}
\newcommand{\quick}{quick fix}

\newcommand{\patternlabel}{EP}

\newcommand{\percentjavaexpt}{72\%}
\newcommand{\projectssurvey}{124}

\newcommand{\emails}{2,000}
\newcommand{\answers}{164}
\newcommand{\panswers}{8.2\%}
\newcommand{\supported}{8}

\newcommand{\psupported}{89\%}
\newcommand{\pullrequests}{16}
\newcommand{\acceptpullrequests}{6}
\newcommand{\totalfeedbackpullrequests}{10}
\newcommand{\acceptpullrequeststext}{six}
\newcommand{\rejectedpullrequeststext}{four}

\newcommand{\percentacceptedPullRequests}{60\%}
\newcommand{\ppullrequests}{50\%}

\newcommand{\projects}{9}
\newcommand{\projectsspell}{nine}
\newcommand{\edits}{89}

\newcommand{\newpatterns}{57}
\newcommand{\percentneepatterns}{64\%}

\newcommand{\maxEdits}{288,899}
\newcommand{\totaleditsinclusters}{205,934}
\newcommand{\maxNumberOfEditsInClusters}{2,706}
\newcommand{\percentEditsInClusterWithMultipleEdits}{71\%}
\newcommand{\maxClusters}{110,384}
\newcommand{\clustersMoreThanTwoEdits}{39,104}

\newcommand{\numbereditpatters}{9}

\newcommand{\RQA}{How effective is \technique{} in identifying quick fixes?\xspace}
\newcommand{\RQB}{Do developers prefer \quickpl{} discovered by \technique{}?\xspace}
\newcommand{\RQC}{Do developers adopt \quickpl{} discovered by \technique{}?\xspace}

\makeatletter
\newcommand\Label[1]{&\refstepcounter{equation}(\theequation)\ltx@label{#1}&}
\makeatother
\usepackage{framed}
   {\endMakeFramed}

\usepackage{fancybox}


\begin{document}
\sloppy
\title{
{Learning Quick Fixes
from Code Repositories}
}


\author{
    \IEEEauthorblockN{Reudismam Rolim\IEEEauthorrefmark{1}, Gustavo Soares\IEEEauthorrefmark{2}, Rohit Gheyi\IEEEauthorrefmark{3}, Titus Barik\IEEEauthorrefmark{2}, Loris D'Antoni\IEEEauthorrefmark{4}}
    \IEEEauthorblockA{\IEEEauthorrefmark{1}UFERSA, Brazil, \IEEEauthorrefmark{2}Microsoft, USA, \IEEEauthorrefmark{3}UFCG, Brazil,  \IEEEauthorrefmark{4}University of Wisconsin-Madison, USA}
    \IEEEauthorblockA{reudismam.sousa@ufersa.edu.br, gsoares@microsoft.com, rohit@dsc.ufcg.edu.br, loris@cs.wisc.edu}
    \vspace*{1.0cm}
}

\maketitle
\thispagestyle{plain}
\pagestyle{plain}

\begin{abstract}
Code analyzers such as Error Prone and FindBugs
detect code patterns symptomatic of bugs, performance issues, or 
bad style. These tools express patterns as \quickpl{}
that detect and rewrite unwanted code.
However, it is difficult to come up with new \quickpl{}  and decide which ones are useful and frequently appear in real code.
We propose to rely on the collective wisdom of programmers and learn \quickpl{} from revision histories in software repositories.
We present \technique{}, a tool for  
discovering common Java edit patterns in code repositories.
Given code repositories and their revision histories, \technique{} 
\rone identifies code edits from revisions  and
\rtwo clusters edits into sets that can be described using an edit pattern.
The designers of code analyzers can then inspect the patterns and add 
the corresponding quick fixes to their tools.
We ran \technique{} on \projectsspell{}  popular
GitHub projects, and it discovered \edits{} useful edit patterns that appeared in 3 or more projects.
Moreover,~\percentneepatterns{} of the discovered patterns did not appear in existing tools.
We then conducted a survey with \answers{} programmers from 
\projectssurvey{} projects and found that programmers significantly preferred eight out of the nine of the discovered patterns. 
Finally, we submitted \pullrequests{} pull requests applying 
our patterns to \projects{} projects and, at the time of the writing, programmers accepted \acceptpullrequests{} (60\%) of them. The results of this work aid toolsmiths in discovering quick fixes and making informed decisions about which quick fixes to prioritize based on patterns programmers actually apply in practice.

\end{abstract}



\section{Introduction}

Programmers often detect code patterns that may lead
to undesired behaviors (e.g., inefficiencies) and
apply simple code edits
to ``fix'' these patterns.
These patterns are often hard to spot
because they depend on the
style and properties of the programming language in use.
Tools such as Error Prone~\cite{ErrorProne}, 
ReSharper~\cite{resharper}, Coverity~\cite{coverity},
FindBugs~\cite{AY08USIN}, PMD~\cite{pmd},
and Checkstyle~\cite{checkstyle} help programmers by
automatically detecting and sometimes removing several suspicious code patterns, 
potential bugs, or instances of bad code style.
For example,
PMD can detect instances where
the method \code{size} is used to check whether a 
list is empty and proposes to replace such instance with the method \code{isEmpty}.
For the majority of collections, these two ways to check emptiness are equivalent, but
for some collections---e.g., \code{ConcurrentSkipListSet}---computing the size of a
list is not a constant-time operation~\cite{URLORACLECONCURRENTSKIPLISTSET}.
We refer to these kinds of edit patterns as \textit{\quickpl{}}.

All the aforementioned tools rely on a predefined catalog of \quickpl{} (usually expressed as rules), each used to detect and potentially fix a pattern. 
These catalogs have to be updated often due 
to the addition of new language features (e.g., new constructs in new versions of Java), new style guidelines, or
simply due to the discovery of new patterns.
However, 
coming up with what edit patterns are useful and common
is a
challenging and time-consuming task that is currently performed in an ad-hoc fashion---i.e.,
new rules for \quickpl{} are added on a as-needed basis.
\begin{result}
The lack of a systematic way of discovering new \quickpl{}
makes it hard for code analyzers to stay up-to-date with the latest code practices and language features.
\end{result}

\begin{figure}[t!]
\begin{subfigure}{0.47\textwidth}
\begin{lstlisting}[language=diff, frame = single, xleftmargin=.03\textwidth]
   //...
-  } else if (args[i].equals("--launchdiag")) {		 
+  } else if ("--launchdiag".equals(args[i])) {
     launchDiag = true;
-  } else if (args[i].equals("--noclasspath") 
-    || args[i].equals("-noclasspath")) {		 
+  } else if ("--noclasspath".equals(args[i]) 
+    || "-noclasspath".equals(args[i])) {
   //...
\end{lstlisting}
\caption{Concrete edits applied to the Apache Ant source code in the project commit history (\url{https://github.com/apache/ant/commit/b7d1e9b}).\label{fig:motivation}}
\vspace{3mm}
\end{subfigure}

\begin{subfigure}{0.5\textwidth}
\centering
\noindent\begin{minipage}{0.45\textwidth}
\begin{lstlisting}[language=trans, frame = single, numbers=none, numbersep=5pt, xrightmargin=0.02\textwidth, breakatwhitespace=false, breaklines, linebackgroundcolor={\lstcolorlines[lightgray!80]{2,2}}]
@BeforeTemplate
boolean b(String v1, 
 StringLiteral v2) {
  return v1.equals(v2);
}
\end{lstlisting}
\end{minipage}
    \hbox{\hspace{0.3em}\includegraphics[width=0.03\textwidth]{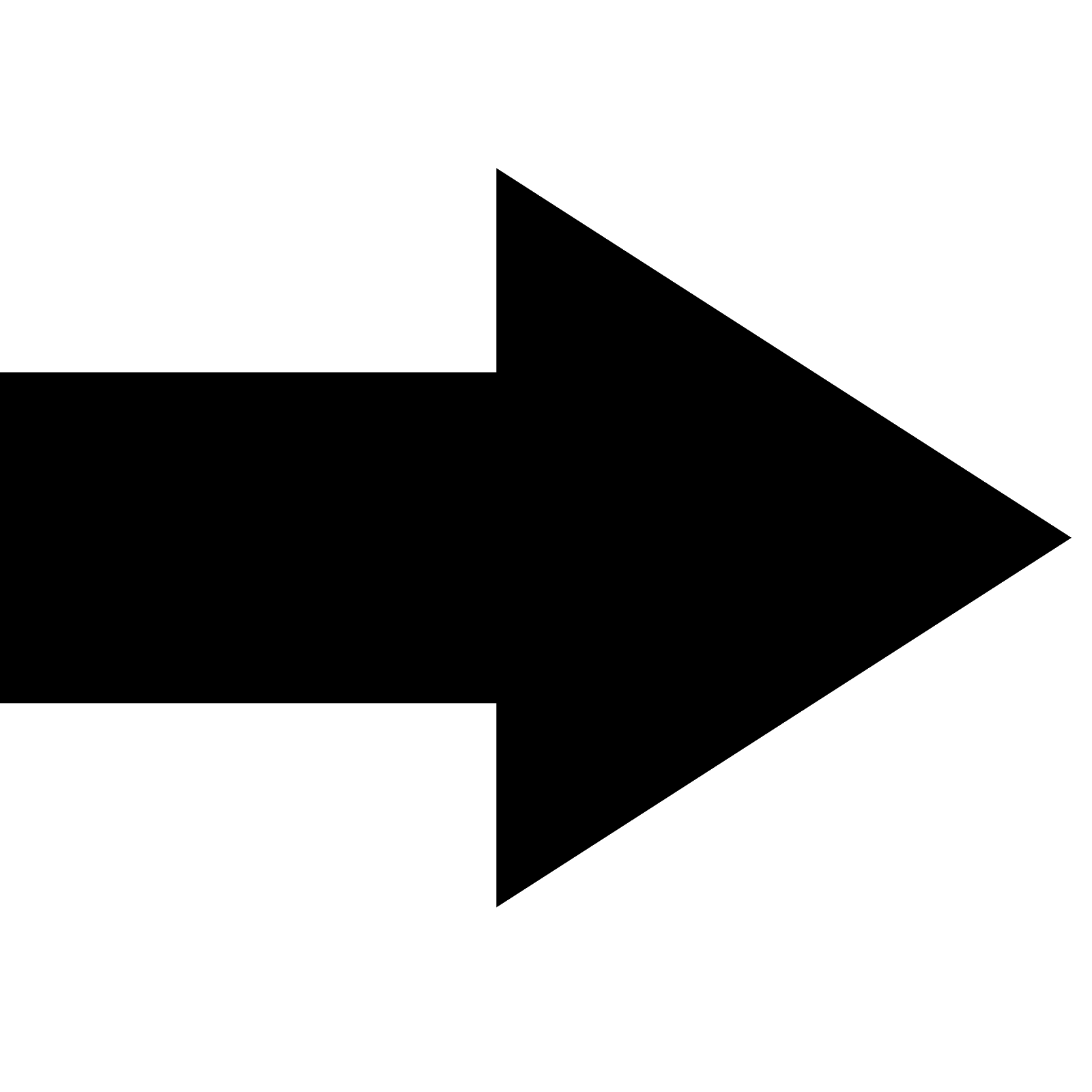}\hspace{-0.3em}}
\begin{minipage}{0.46\textwidth}
\begin{lstlisting}[language=trans, frame = single, numbers=none, numbersep=5pt, xleftmargin=0.05\textwidth, breakatwhitespace=false, breaklines, linebackgroundcolor={\lstcolorlines[lightgray!80]{2,2}}]
@AfterTemplate
boolean a(String v1, 
 StringLiteral v2) {
  return v2.equals(v1);
}
\end{lstlisting}
\end{minipage}
\caption{Abstract quick fix in Refaster-like syntax for edits in Figure~\ref{fig:motivation}.}
\label{fig:equalstransformation}
\end{subfigure}
\caption{\technique{} (a) mines concrete edits from the code repository history in a project and (b) discovers abstract quick fixes from these edits.}
\end{figure}

For example, consider the edit pattern applied to
the code in Figure~\ref{fig:motivation}.
The edit was performed in the Apache Ant source code.
In the presented pattern, the original code contains three expressions of the form \code{x.equals("str")} 
that compare a variable
 \code{x} of type string to a string literal \code{"str"}. 
Since the variable \code{x} may contain a \code{null} value, 
evaluating this expression may cause a 
\code{NullPointerException}. 
In this particular revision,
a programmer from this project addresses the issue
by exchanging the order of the arguments of the \code{equals} method---i.e., 
by calling the method on the string literal.
This edit fixes the issue since the \code{equals} method checks 
whether the
parameter is \code{null}.
This edit is common and we discovered it occurs in three industrial 
open source projects across GitHub repositories: Apache Ant, Apache Hive, and Google ExoPlayer.
Given that the pattern appears in such large repositories, it makes sense to assume that
it
could also be useful to other programmers
who may not know about it.
Despite its usefulness, a quick fix rule for this edit pattern is not included in the catalog of 
largely used code analysis tools, such as FindBugs, and PMD.
Remarkably, even though the edit is applied in Google repositories, this
pattern does not appear in Error Prone, a code analyzer developed by Google that is internally 
run on the Google's code base.



\textbf{Key insight}
Our key insight is that we can ``discover'' useful patterns and \quickpl{} by
observing how programmers modify code
in
real repositories with large user bases.
In particular, we postulate that an edit pattern that is performed by many programmers across many projects 
is likely to reveal a good \quick{}.


\textbf{Our technique} In this work, we propose \technique{}, a technique for
automatically discovering common Java code edit patterns in
online code repositories. Given code repositories as input, 
\technique{}  
identifies simple edit patterns by comparing 
consecutive revisions in the revision histories.
The most common edit patterns---i.e., those performed across multiple projects---can 
then be inspected to detect
useful ones and add the corresponding \quickpl{} to code analyzers.
For example, 
\technique{} was able to \emph{automatically} analyze the concrete edits
in Figure~\ref{fig:motivation} and generate the
\quick{} in Figure~\ref{fig:equalstransformation}.
We also sent pull requests applying this \quick{} to other parts of the code
in the Apache Ant and Google ExoPlayer projects, and
these pull requests were accepted.

Since we want to detect patterns appearing 
across projects and revisions, \technique{} has to 
analyze large amounts of code, a task that
requires efficient and precise algorithms.
\technique{} focuses on edits performed to individual code locations and uses GumTree~\cite{FA14FINE}, a tree edit distance algorithm,
to efficiently
extract concrete abstract-syntax-tree edits from pairs of revisions---i.e., 
sequences of tree operations such as insert, delete, and update.
\technique{} then
uses a greedy algorithm to detect subsets of concrete edits that can
be described using the same edit pattern.
To perform this last task efficiently, \technique{}
uses a variant of a technique called anti-unification~\cite{KU14ANTI}, which
is commonly used in inductive logic programming.
Given a set of concrete edits, the anti-unification algorithm finds the least general generalization of the two edits---i.e., the largest
pattern shared by the edits.

\vspace{2mm}\noindent\textbf{Contributions} This paper makes the following contributions:

\begin{itemize}
    \item \technique{}, an automatic 
     technique for discovering common edit patterns
    in large online repositories.
    \technique{} uses concepts such as
    AST edits and $d$-caps, and it 
    applies the technique of anti-unification from inductive logic programming
    to the problem of mining edit patterns
    (\S~\ref{s:technique}).
    \item A mixed-methods evaluation of the effectiveness of \technique{} at discovering \quickpl{} and the
    quality of the discovered \quickpl{} (\S~\ref{sec:methodology}~and~\ref{sec:results}).
    When ran on \projectsspell{}  popular
    GitHub projects, \technique{}  discovered \edits{} edit patterns that appeared in 3 or more projects.
    Moreover,~\percentneepatterns{} of the discovered patterns did not appear in existing tools.
    Through a survey on a subset of the discovered patterns, 
    we showed that programmers significantly preferred \supported{}/9 (\psupported{}) of our patterns.  
    Finally, programmers 
    accepted \acceptpullrequests/\totalfeedbackpullrequests{} (\percentacceptedPullRequests{}) pull requests
    applying our patterns to their projects, showing that developers are willing to apply our patterns to their code.
\end{itemize}

\section{\technique{}}
\label{s:technique}

\begin{figure}
    \centering
    \includegraphics[width=0.47\textwidth]{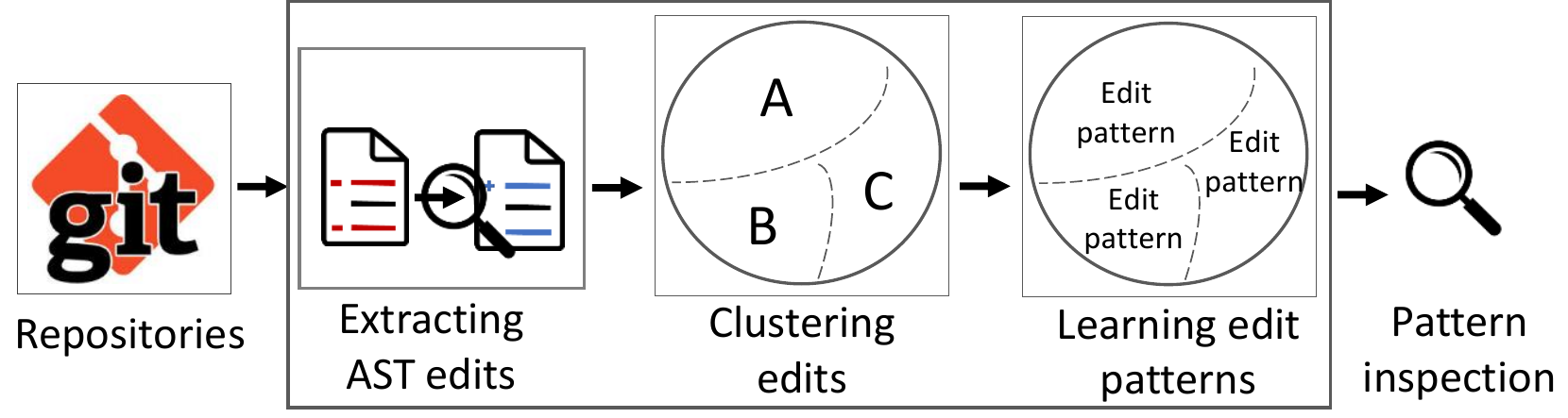}
    \caption{\technique{}'s work-flow.}
    \label{fig:process}
\end{figure}

We now describe \technique{}, our technique for 
automatically discovering common edit patterns
in code repositories. Given 
repositories as input, \technique{}:
\rone identifies concrete code edits by comparing pairs 
of consecutive revisions (\S~\ref{s:edits}),
\rtwo clusters edits into sets that can be described using the same 
edit pattern
and learns an edit pattern for each cluster (\S~\ref{s:template}~and~\ref{s:clustering}). 
Figure~\ref{fig:process} shows the work-flow of \technique{}.

\subsection{Extracting concrete AST edits}
\label{s:edits}

The initial input of \technique{} is a set 
 $revs=\{R_1,\ldots, R_n\}$ where
each $R_i$ is a revision history $r_1 r_2 \cdots r_k$  from a different project (i.e., a sequence of revisions).
For each pair $(r_i, r_{i+1})$ of consecutive revisions, 
\technique{} 
analyzes the differences between the Abstract Syntax Trees (ASTs)\footnote{\technique{} uses Eclipse JDT~\cite{eclipsejdt} to extract
partial type annotations of the ASTs. In our implementation, we use these type
annotations to create a richer AST with type information---i.e., every node
has a child describing its type. For simplicity,
we omit this detail in the rest of the section.
}
of $r_i$ and $r_{i+1}$
and uses a Tree Edit Distance (TED) algorithm to identify a set of tree edits $\{e_1, e_2, ... , e_l\}$ that
when applied to the AST $t_i$ corresponding to $r_i$ yields the AST $t_{i+1}$ corresponding to $r_{i+1}$. 
In our setting, an edit is one of the following:

\begin{LaTeXdescription}
	\item[insert(x, p, k):] insert a leaf node $x$ as $k^{th}$ child of parent node $p$.
	The current children at positions $\geq k$ are shifted one position to the right.
	\item[delete(x, p, k):]
	delete a leaf 
	node $x$ which is the $k^{th}$ child of parent node $p$. The deletion may cause new nodes to become leaves when all
	their children are deleted. Therefore we can delete a whole tree through repeated 
	bottom-up deletions.
	\item[update(x, w):] replace a leaf 
	node $x$ by a leaf node $w$.
	\item[move(x, p, k):] move tree $x$ 
	to be the $k^{th}$ child of parent node $p$. 
	The current children at positions $\geq k$ are shifted one position to the right.
\end{LaTeXdescription}
Given a tree $t$, let $s(t)$ be the set of nodes in the tree $t$.
Intuitively, solving the TED problem amounts to identifying a  partial
mapping $M:s(t)\mapsto s(t')$ between source tree $t$ and target tree $t'$ nodes.
The mapping can then be used to detect which nodes are preserved
by the edit and in which positions they appear.
When a node $n\in s(t)$ is not mapped to any $n'\in s(t')$, then $n$ was deleted.
%

\begin{figure}[H]
\centering
\begin{subfigure}{0.21\textwidth}
\centering
\frame{\includegraphics[width=0.97\textwidth]{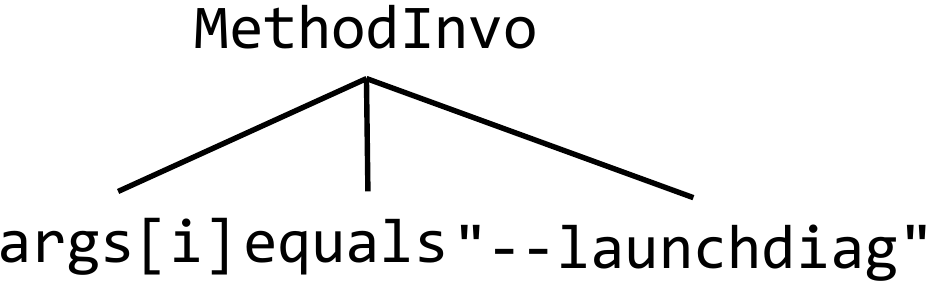}}
\end{subfigure}
\includegraphics[width=0.03\textwidth]{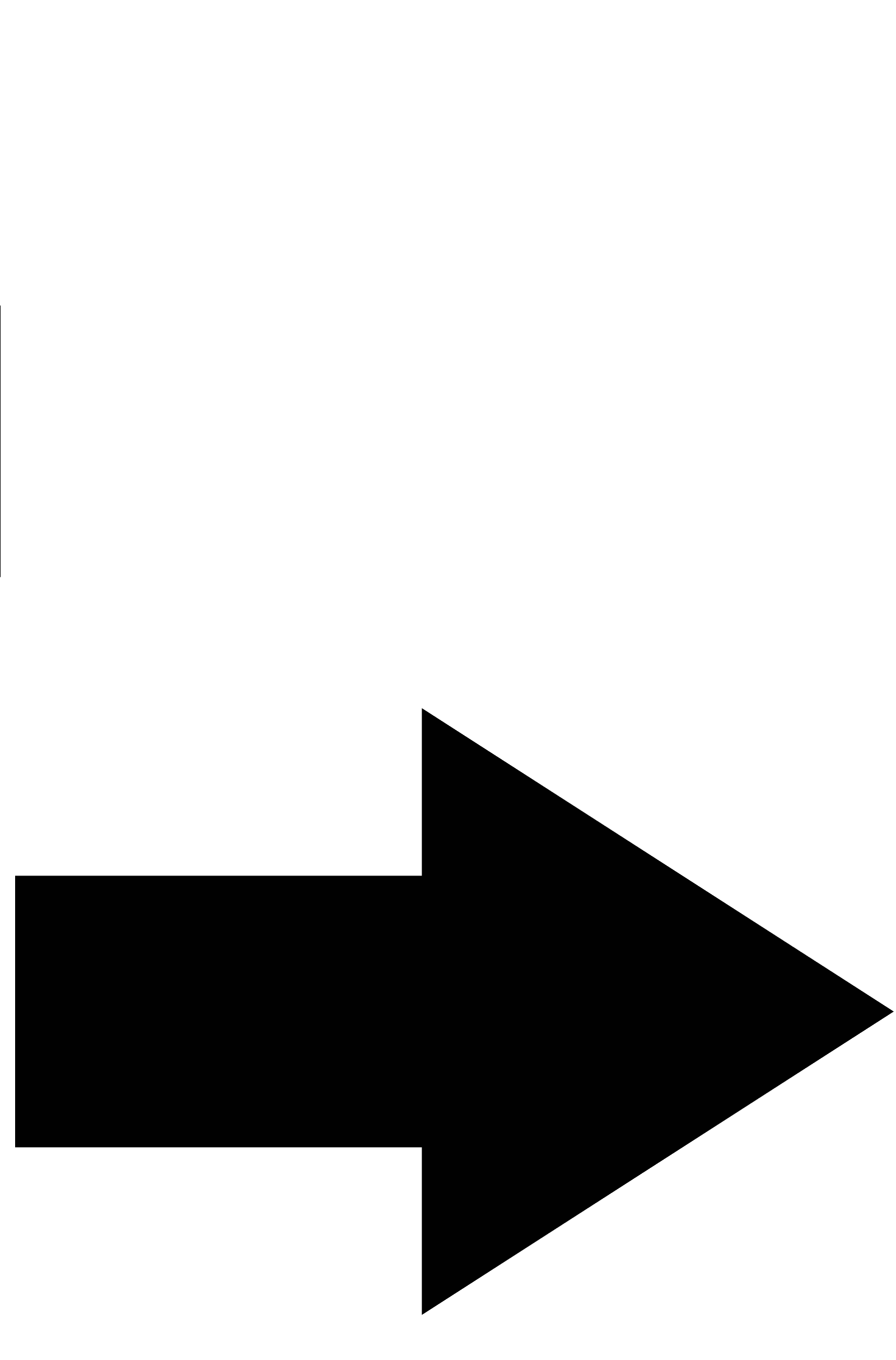}
\begin{subfigure}{0.21\textwidth}
\centering
\frame{\includegraphics[width=0.97\textwidth]{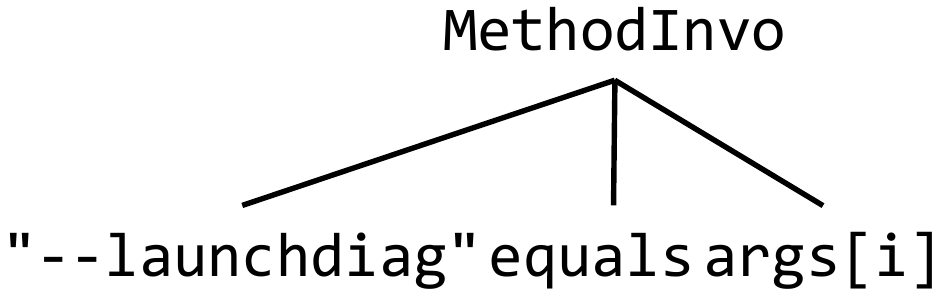}}
\end{subfigure}
\caption{Before-after version for the first edited line of code.}
\label{fig:tree}
\end{figure}

Of the many existing tools that are available for computing tree edits over Java source code,
\technique{} builds on GumTree~\cite{FA14FINE}, a tool that 
focuses on finding edits that are representative of those intended by programmers instead 
of just finding the smallest possible set of tree edits.
To give an example of edits computed by GumTree, let's look at 
the first two lines in Figure~\ref{fig:motivation}.  
Figure~\ref{fig:tree} illustrates the ASTs  corresponding to 
\code{args[i].equals("--launchdiag")} and
\code{"--launchdiag".equals(args[i])}, respectively. 

To produce the modified version of the code in line 2 at Figure~\ref{fig:motivation}, 
GumTree learned four edits:
\begin{LaTeXdescription}
    \item \texttt{insert("--launchdiag", MethodInvo, 0)}
    \item \texttt{insert(equals, MethodInvo, 1)}
    \item \texttt{delete(equals, MethodInvo, 3)}
    \item \texttt{delete("--launchdiag", MethodInvo, 3)}
\end{LaTeXdescription}
These edits move the string literal \code{"--launchdiag"} and the 
\code{equals} node so that they appear in front of \code{args[i]}. 

For our purposes, the edits computed by GumTree are at too low granularity since
they modify nodes instead of expressions.
In particular, we are interested in  detecting edits to entire subtrees---e.g.,
an edit to a 
method invocation \code{args[i].equals("--launchdiag")} instead of to individual
nodes inside it.
\technique{} identifies subtree-level edits
by 
grouping edits that belong to the same connected components.
Concretely, \technique{} identifies connected components by analyzing the parent-child and sibling relationships between the nodes appearing in the tree edits. 
Two edits $e_1$ and $e_2$ belong to the same component if
\rone the two nodes $x_1$ and $x_2$ modified by
$e_1$ and $e_2$ have the same parent, or 
\rtwo the $x_1$ (resp. $x_2$) is the parent of $x_2$ (resp. $x_1$). 
For instance, the previously shown edits are associated to two nodes $x_1 = $ \code{"--launchdiag"} and $x_2 = $ \code{equals}. These nodes are connected since
they have the same parent node---i.e., the method invocation.

Once the connected components are identified, \technique{} 
can use them to identify tree-to-tree 
mappings between subtrees inside the original and modified trees like the one showed in Figure~\ref{fig:tree}. We call this mapping a \textit{concrete edit}. A concrete edit is a pair $(i, o)$ consisting of two components
\rone the tree $i$ in the original version of the program, and
\rtwo the tree $o$ in the modified version of the program. 
This last step completes the first phase of our algorithm,
which, given a set of revisions $\{R_1, \ldots, R_n\}$, outputs
a set  of concrete edits $\{(i_1,o_1), \ldots, (i_k,o_k)\}$.

\subsection{Learning edit patterns}
\label{s:template}
Once \technique{} has identified concrete edits---i.e., 
pairs of trees $\{(i_1,o_1)\ldots(i_n,o_n)\}$---it tries to 
group ``similar'' concrete edits and to
generate an edit pattern consistent with all the edits in each group.
An \textit{edit pattern}
is a rule $r=\tau_i \mapsto \tau_o$
with two components:
\rone the template $\tau_i$, which is used to decide whether a subtree $t$ in the code can be transformed using the rule $r$,
\rtwo the template $\tau_o$, which describes how the tree matching $\tau_i$ should be transformed by $r$.
In this section, we focus on computing $r$ from a set of concrete edits $\{(i_1,o_1)\ldots(i_n,o_n)\}$---i.e.,
we assume we are given
a group of concrete edits that can be described by the same edit pattern. We will discuss our clustering algorithm for creating the groups in the next section.

\begin{figure}[!htbp]
    \centering
    \includegraphics[width=\columnwidth]{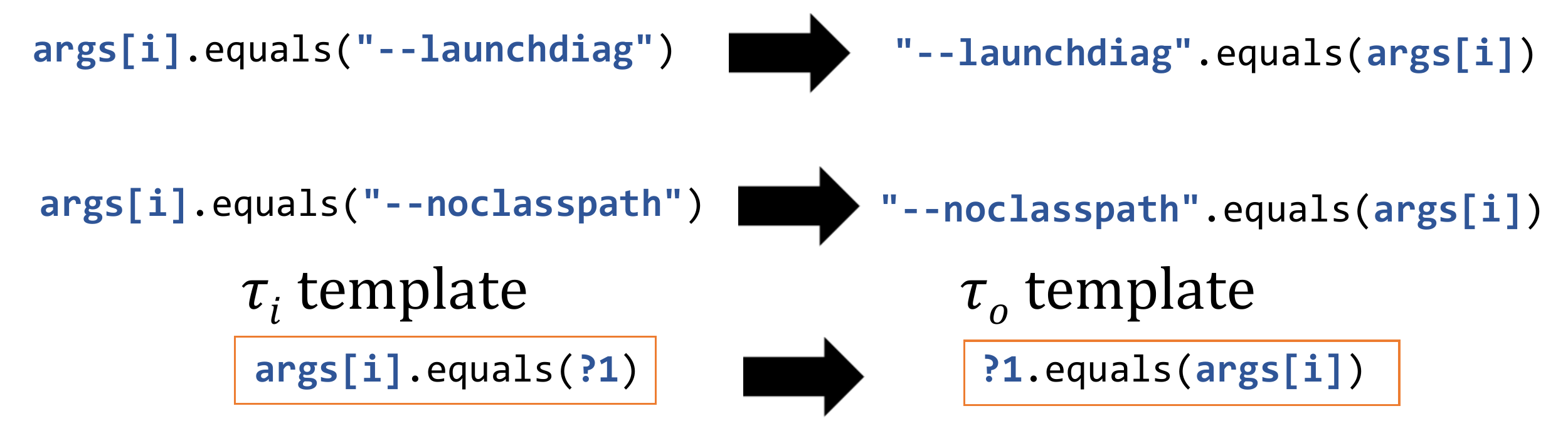}
    \caption{Concrete edits and their input-output templates.}
    \label{fig:template}
\end{figure}

A template $\tau$ is an AST where leaves can also be holes (variables)
and a tree $t$ matches the template $\tau$ if there exists a 
way to assign concrete values to the holes
in $\tau$ and obtain $t$---denoted $t\in L(\tau)$.
Given a template $\tau$ over a set of holes $H$, 
we use $\alpha$ to denote a substitution from $H$ to concrete trees and
$\alpha(\tau)$ to denote the application of the substitution $\alpha$ to the holes in $\tau$.  
Figure~\ref{fig:template} shows the first
two concrete edits from Figure~\ref{fig:motivation}
and the templates $\tau_i$ and $\tau_o$ describing the edit pattern obtained
from these examples.
Here, the template $\tau_i$ matches any expression calling the method
\code{equals} with first argument \code{args[i]}
 and any possible second argument. 
The two substitutions $\alpha_1$=$\{?_1$=\code{"--launchdiag"}$\}$ and
$\alpha_2$=$\{?_1$=\code{"--noclasspath"}$\}$
yield  the expressions $\alpha_1(\tau_i)=$\code{args[i].equals("--launchdiag")} and 
$\alpha_2(\tau_i)=$\code{args[i].equals("--noclasspath")}, respectively.
 The template $\tau_o$ is similar to $\tau_i$ and note
 that the hole $?_1$ appearing in $\tau_o$ is the same as the one appearing in $\tau_i$.

\begin{definition}
Given a set of concrete edits $S=\{(i_1,o_1),\ldots,(i_n,o_n)\}$,
an edit pattern $r=\tau_i\mapsto \tau_o$,
is consistent with $S$ if:
\rone 
the set of holes in $\tau_o$ is a subset of the set the holes appearing in $\tau_i$
\rtwo for every $(i_k,o_k)$ there exists a substitution 
$\alpha_k$ such that
$\alpha_k(\tau_i)=i_k$ and $\alpha_k(\tau_o)=o_k$.
\end{definition}



In the rest of the section, we describe how we obtain the edit pattern
from the concrete edits.
We start by describing how the input template $\tau_i$ is computed.
Our goal is to compute a template $\tau_i$ such that every AST $i_j$ can match the template $\tau_i$---i.e., $i_j\in L(\tau_i)$.
In general, the same set of ASTs can be matched by multiple different templates,
which could contain different numbers of nodes and holes.
Typically, a template with more concrete nodes and fewer holes
is more precise---i.e., will match fewer concrete ASTs---whereas a 
template with few concrete nodes will be more general---e.g.,
\code{$?_1$.equals($?_2$)} is more general than \code{arg[i].equals($?_1$)}.
Among the possible templates, we want the \textit{least general template}, which preserves the maximum common 
nodes for a given set of trees. 
The idea is to preserve the maximum amount of
shared information between the concrete edits.
Even when an edit is too specific,
 we will obtained the desired template when provided
 with appropriate concrete edits.
 In our running example, when encountering an expression of the form
 \code{x.equals("abc")}, we will obtain the desired, more general template \code{$?_1$.equals($?_2$)}.

Remarkably, the problem we just described is tightly related
to the notion of anti-unification used in
logic programming~\cite{BA17UNRA, BA17HIGH}.
Given two trees $t_1$ and $t_2$, 
the anti-unification algorithm produces
the least general template 
$\tau$ for which there exist substitutions $\alpha_1$ and $\alpha_2$ 
such that $\alpha_1(\tau)$ = $t_1$ and $\alpha_2(\tau)$ = $t_2$. 
In our tool we use the implementation of anti-unification from Baumgartner et al.~\cite{BA17HIGH},
which runs in linear time.
Using this algorithm, we can generate the least general templates
$\tau_i$ and $\tau_o$
that are consistent with the input and output trees in the 
concrete edits.
For now, the two templates will have distinct sets of holes, but each template can contain the same
hole in multiple locations.

\begin{figure}[!htbp]
    \centering
    \includegraphics[width=\columnwidth]{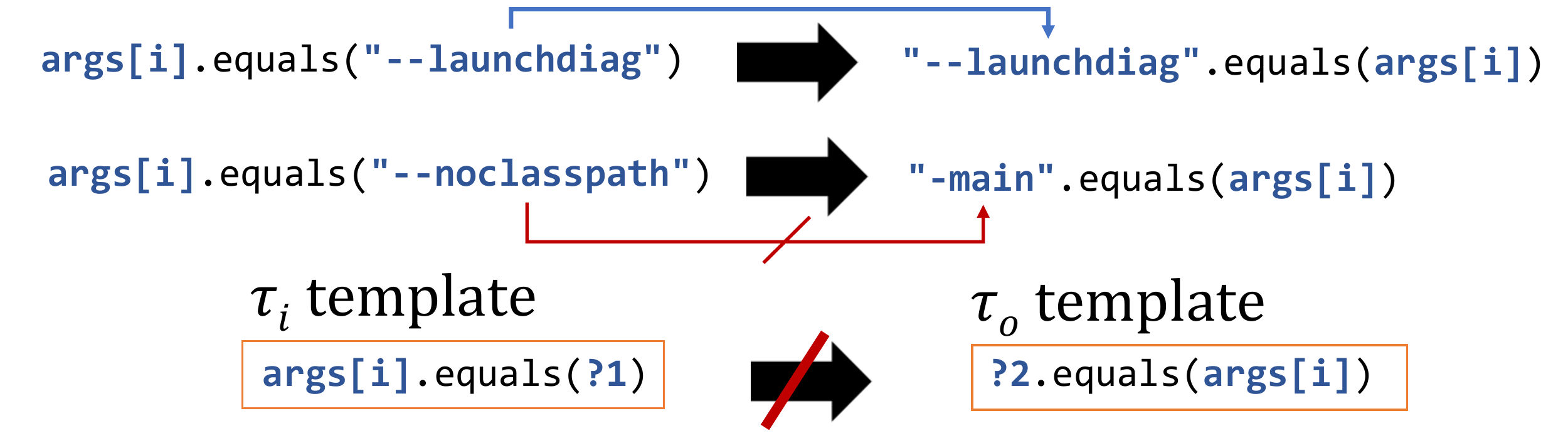}
    \caption{Incompatible concrete edits.}
    \label{fig:invalidrule}
\end{figure}

At this point, we have the template for the inputs $\tau_i$ and the template for the output $\tau_o$.
However, \technique{} needs to analyze whether these templates describe an edit pattern $r = \tau_i\mapsto\tau_o$---i.e., whether
there is a way to map the holes of $\tau_i$ to the ones of $\tau_o$.
This mapping can be computed by finding, for every hole $?_2$ in $\tau_o$,
a hole $?_1$ in $\tau_i$ that applies the same substitution with respect
to all the concrete edits.
To illustrate a case where finding a mapping is not possible, let's look at Figure~\ref{fig:invalidrule}. Although we can learn templates $\tau_i$ and $\tau_o$, these templates
cannot describe an edit pattern since it is impossible to come up with a mapping between the holes of the two templates that is consistent
with all the substitutions. 
In this case, the substitution
for $?_1$ in $\tau_i$ is incompatible with the substitution
for $?_2$ in $\tau_o$ because \code{"--noclasspath"} is mapped to \code{"-main"}, but the content of these substituting trees differs.
In addition, in our implementation, we avoid to group concrete edits that are compatible, accordingly to our definition but apply to different methods. For instance, the concrete edits (\code{args[i].equals("--launchdiag")}, \code{"--launchdiag".equals(args[i]))} and (\code{args[i].equalsIgnoreCase("--launchdiag")}, \code{"--launchdiag".equalsIgnoreCase(args[i]))}  should not be in the same cluster since they are applied to the \code{equals} and \code{equalsIgnoreCase} methods, respectively.

Since the templates are the least general, if no mapping between the holes exists, there exists
no edit pattern consistent with the concrete edits---i.e., our algorithm, given a set of edits, finds a rule in our format consistent with the edits if and only if one exists. 
Therefore, \technique{} finds all correct rules in our format and does not miss potential ones.

\begin{theorem}[Soundness and Completeness]
Given a set of concrete edits $S=\{(i_1,o_1),\ldots,(i_n,o_n)\}$,
\technique{} returns 
an edit pattern $r=\tau_i\mapsto \tau_o$ consistent with $S$
if and only if some edit pattern $r'=\tau_i'\mapsto \tau_o'$ consistent with $S$ exists.
\end{theorem}

\subsection{Clustering concrete edits}
\label{s:clustering}
In this section, we show how \technique{}
groups concrete edits into clusters that share the same
edit pattern.
\technique{}'s clustering algorithm receives a set of concrete edits $\{(i_1,o_1)\ldots(i_n,o_n)\}$ and 
uses a greedy approach. 
The clustering algorithm
starts with an empty set of clusters.
Then, for each concrete edit $(i_k, o_k)$ 
and for each cluster $c$, \technique{}
checks, using the algorithm from Section~\ref{s:template}, if adding $(i_k, o_k)$ 
to the concrete edits of cluster $c$ gives an edit pattern. 
When this happens, the cluster $c$ is added to a set of
cluster candidates and the cost of adding $(i_k, o_k)$ to $c$ is computed.
\technique{} then adds $(i_k, o_k)$ to candidate cluster of
minimum cost, or it creates
a new cluster with just $(i_k, o_k)$
if no candidate exists.
The complexity of this algorithm is $O(n^2)$ where $n$ is the number of edits since, for each edit, we have to 
search  which cluster
the edit should be included in. 

When multiple clusters can receive a new concrete edit, we use the following cost function to decide which cluster to add the edit to.
The cost of
adding an edit $(i_k, o_k)$ to cluster $c$ with corresponding template $\tau$ is computed as follows. 
First, \technique{} anti-unifies
$\tau$  and the tree in the concrete edit we are trying to cluster.
Let  $\alpha_k$ be the substitution for the result of the anti-unification
and let $\alpha_k(?_i)$ be the tree substituting hole $?_i$.
We define the size of a tree as the number of leaf nodes 
inside it, which intuitively captures the 
number of names and constants 
in the AST. 
We denote the cost of an anti-unification  as the sum of 
the sizes of each $\alpha_k(?_i)$
minus the total number of 
holes (the same metric proposed by Bulychev et al.~\cite{BU09ANEV} in the context of clone detection). 
Intuitively, we want the sizes of substitutions to be small---i.e., we
prefer more specific templates.
For instance, assume we have a cluster consisting of a single tree
\code{args[i].equals("--launchdiag")}.
Upon receiving a new tree \code{args[i].equals("--noclasspath")},
anti-unifying the two trees yields the template \code{args[i].equals($?_1$)} with 
substitutions $\alpha_1$=$\{?_1$=\code{"--launchdiag"}$\}$ and
$\alpha_2$=$\{?_1$=\code{"--noclasspath"}$\}$.
We have concrete nodes 
\code{"--launchdiag"} and \code{"--noclasspath"} each of size one,
and a single hole $?_1$.
The cost will be $2 - 1 = 1$.
The final cost of adding an edit to a cluster is the sum of the cost to anti-unify
$\tau_i$ and $i_k$
and the cost to anti-unify $\tau_o$ and $o_k$.



\textit{Predicting promising clusters}
For large repositories,
the total number of concrete edits may be huge and it will be 
unfeasible to compare all edits to compute the clusters.
To address this problem,
\technique{} only clusters concrete edits which are 
``likely''
to produce an edit pattern.
In particular, \technique{} uses \textit{d-cap}s~\cite{EV07CLON, BA17UNRA, AN13ASTU}, a technique for
 identifying repetitive edits.
Given a number $d \geq 1$ and a template $\tau$, a $d$-cap
is a tree-like structure 
obtained by replacing all subtrees of depth $d$ and left nodes in the template $\tau$
with holes. 
The $d$-cap works as a hash-index for 
sets of potential clusters. 
For instance, let's look at the left-hand side of Figure~\ref{fig:tree}, 
which is the tree representation 
of the node \code{args[i].equals("--launchdiag")}. A 1-cap replaces 
all the nodes at depth one with holes. For our example, \code{args[i]},
\code{equals}, and \code{"--launchdiag"} will be replaced with holes, outputting
the $d$-cap \code{$?_1$.$?_2$($?_3$)}.
\technique{} uses the $d$-caps as a pre-step\iffull\else\:in the clustering algorithm\fi.
For all concrete edits with the same $d$-cap for the input tree $i_k$ and for the output tree $o_k$, 
\technique{} uses the clustering algorithm described in
Section~\ref{s:clustering} to compute the clusters for 
all concrete edits in these $d$-cap.
This heuristic makes our clustering algorithm practical as it avoids considering all
example combinations. However, it also comes at the cost of sacrificing completeness---i.e., 
two concrete edits for which there is a common edit pattern might be placed
in different clusters. 

\section{Methodology}
\label{sec:methodology}

\cref{sec:rqs} states our research questions. \cref{sec:methodology:technique} describes the evaluation for the \technique{} technique. \cref{sec:methodology:survey} describes a survey study in which programmers evaluate quick fixes discovered by \technique{}. \cref{s:pullrequest} describes a validation study in which we submit these quick fixes as pull requests to GitHub projects.

\subsection{Research Questions}
\label{sec:rqs}

We investigate the following three research questions:

\begin{enumerate}[leftmargin=3em,label=\textbf{RQ\arabic*}]
\item \RQA 
\item \RQB
\item \RQC
\end{enumerate}

The answer to the first research question characterizes the edit patterns that \technique{} is able to discover, and whether these edit patterns can be framed in terms of existing code analysis tool rulesets. The answers to the second and third research questions address whether these identified edit patterns are useful for developers, through different perspectives: preference (RQ2) and adoption (RQ3).

\subsection{Evaluation Methodology for \technique{}}
\label{sec:methodology:technique}

\textbf{Data collection} We selected \projects{} popular GitHub Java projects (Table~\ref{t:projects}) from a list of previously studied projects~\cite{SI16WHYW}.
\changes{The project selection influences quantity and quality of the discovered patterns. We select mature popular projects that have long history of edits, experienced developers who detect problems during code reviews, and several collaborators (avg. 89.11) with different levels of expertise (low expertise collaborators likely submit pull-requests that need quick fixes). Analyzing many projects is prohibitive given our resources, thus we selected a subset of projects with various sizes/domains.} We favored projects containing between 1,000 and 15,000 revisions, to have a sample large enough to identify many patterns but not too big due to the time required to evaluate all revisions. The projects have size ranging from 24,753 to 1,119,579 lines of code. In total, the sample contains 43,113 revisions, which were input into \technique{}.


\textbf{Benchmarks} \technique{} found~\maxEdits{} single-location edits which were clustered in~\maxClusters{} clusters. Of these clusters,~\clustersMoreThanTwoEdits{} contained more than one edit---i.e., \technique{} could generalize multiple examples to a single \emph{edit pattern}. 
The~\clustersMoreThanTwoEdits{} edit patterns covered \totaleditsinclusters{}/\maxEdits{} single-location edits (\percentEditsInClusterWithMultipleEdits{}).
The distribution of these edit patterns is reminiscent of a long-tail one: the most-common edit pattern having~\maxNumberOfEditsInClusters{} concrete edits,
$0.06\%$ of the edit patterns cover 10\% of the concrete edits, and 5.3\% of the edit patterns cover 20\% of the concrete edits.


We performed the experiments on a PC running Windows 10 with a processor Core i7 and 16GB of RAM. 
We obtained the revision histories of each repository
using JGit~\cite{jgit}. 
\technique{} took 5 days to analyze the \projects{} projects (approximately 10 seconds per commit). Most of the time is spent checking out revisions,
a process that can be done incrementally for future projects.

\textbf{Edit pattern sampling} 
To facilitate manual investigation of these edit patterns, we empirically identified a ``Goldilocks'' cut-offs for edit patterns found in $n$ or more projects that would allow us to practically inspect each patterns (at $n \ge 1$: 110,384 edit patterns, 2: 1,759, 3: 493, 4: 196; 5: 89, 6: 47, 7: 19, 8: 6, 9: 1). From this distribution, we choose the 493 patterns found in 3 or more projects as a reasonable cut-off for subsequent analysis.

\begin{table}[!t]
\centering
\caption{Projects used to detect edit patterns\label{t:projects}}
\begin{tabular}{lrrr}
\toprule
\textbf{Project} & \textbf{Edits} & \textbf{LOC} & \textbf{Revisions}\\
\midrule
Hive & 94,921  & 1,119,579  &  11,467      \\
Ant        & 49,680  & 137,203    &  13,790      \\
Guava      & 28,784  & 325,902    &  4,633       \\
Drill      & 26,173  & 350,756    &  2,902       \\
ExoPlayer  & 20,726  & 85,305     &  3,875       \\
Giraph     & 8,836   & 99,274     &  1,062       \\
Gson       & 4,435   & 24,753     &  1,393       \\
Truth      & 3,857   & 27,427     &  1,137       \\
Error Prone & 3,200   & 116,023    &  2,854       \\ \midrule
Totals     & 240,612 & 2,286,222  &  43,113      \\ \bottomrule
\end{tabular}
\end{table}

\textbf{Analysis} \emph{Phase I---Spurious pattern elimination} To assess the effectiveness of \technique{}, we conducted a filtering exercise to discard spurious edit patterns. Specifically, we discarded edit patterns involving renaming operations---e.g., renaming the variable \texttt{obj} to \texttt{object}---and edit patterns in which none of the authors could identify a logical rationale behind the edit, typically because the patterns were part of some broader edit sequence---e.g., changing \texttt{return true} to \texttt{return null}. 
We eliminated \noise{} spurious patterns, where 61 of them were renaming operation and, for the rest of them, we could not identify a logically meaningful pattern.

\emph{Phase II---Merging duplicate edit patterns} Next, we merged edit patterns that represented the same logical quick fix. To do so, we employed a technique of \emph{negotiated agreement}~\cite{Campbell2013} in which the second and fourth authors collaboratively identified and discarded logically duplicate edit patterns within the sample. 
When there was disagreement about whether two edit patterns were logical duplicates, the authors opted to merge these duplicates, thus penalizing the effectiveness of \technique{}. In other words, this measure is an \emph{upper bound} of the number of duplicates within the edit patterns. 
We merged \duplicates{} duplicated patterns into 17 other patterns.

\emph{Phase III---Cataloging edit patterns} Finally, we classified each of the remaining 
493-\noise{}-\duplicates{}=\edits{} patterns against the eight Java rule-set from the PMD code analyzer tool (for example, ``Performance'' and ``Code Style''). The full rule-set is shown in Table~\ref{t:rulecategories}.
We targeted the PMD rule-set because the PMD developers employed a principled process for designing this rule-set. In particular, a goal of the rule-set was to make the rule-set useful for reporting by third-party tools and techniques.\footnote{\url{https://github.com/pmd/pmd/wiki/Rule-Categories}} 
The first and third authors independently mapped the edit patterns to one of the PMD rule-sets. We computed Cohen's Kappa to assess the measure of agreement, and deferred to the first author's judgment to reconcile disagreement. Finally, the first author used online resources, such as Stack Overflow and documentation from different code analysis tools to tag each quick fix as being available or not available in
the catalog of an existing tool.




\begin{table}[!t]
\centering
\caption{Description of the categories and frequency\label{t:rulecategories}}
\begin{tabular}{lp{4.5cm}r}
\toprule
\textbf{PMD ruleset}       & \textbf{Description}                                                                      & \textbf{$n$} \\
\midrule
Best Practices & Rules which enforce generally accepted best practices.                                             &    19   \\
Code Style     & Rules which enforce a specific coding style.                                                       &    16   \\
Design         & Rules which help you discover design issues.                                                       &    22   \\
Documentation  & Rules which are related to code documentation.                                                     &    5   \\
Error Prone    & Rules which detect constructs that are either broken, extremely confusing or prone to runtime errors. &   12    \\
Multithreading & Rules which flag issues when dealing with multiple threads of execution.                           &      2 \\
Performance    & Rules which flag suboptimal code.                                                                  &     13  \\
Security       & Rules which flag potential security flaws.                                                         &   0 \\ 
\midrule
\textbf{Total} & & \edits{}\\
\bottomrule
\end{tabular}
\end{table}
\subsection{Developer Survey}
\label{sec:methodology:survey}









We conducted a survey to assess programmers' judgments about a subset of our discovered edit patterns.

\textbf{Participants} We randomly invited \emails{} programmers to participate in our survey through e-mail. We obtained these e-mail addresses from author metadata in the commit histories from \projectssurvey{} popular and well-known GitHub Java-based projects~\cite{SI16WHYW}, such as those from Apache, Google, Facebook, Netflix, and JetBrains. We received \answers{} responses (response rate \panswers{}). Through demographic questions in the survey, 118 participants (\percentjavaexpt{}) self-reported having more than five years of experience with Java. Participants also self-reported using tools to flag code patterns, including IntelliJ (72\%), Checkstyle (50\%), Sonar (50\%), FindBugs (43\%), PMD (31\%), Eclipse (8\%), Error Prone (7\%), and others (8\%). Participants did not receive compensation for their responses.

\textbf{Survey protocol} To constrain the survey response time to 5-10 minutes, we presented programmers with \numbereditpatters{} out of the \edits{} edit patterns
using purposive sampling. In other words, we deliberately selected from a design space of candidate edit patterns to balance patterns found in existing analysis tools versus new edit patterns identified in our study. This selection allowed us to verify whether programmers chose patterns found in existing tools as well as to assess their preferences for edit patterns \emph{not} found in current tools.

We presented edit patterns as side-by-side panes (adjacent left and right panes), with one pane having the baseline code pattern (``expected bad'') and the other with the quick fix version of the code pattern (``expected good''). Each pair was randomized and labeled simply as pattern A and pattern B, so that programmers could not obviously identify the quick fix version of the pattern. To assess if developers preferred the quick fix version of the pattern, we presented a five-point Likert-type item scale: strongly prefer (A), prefer (A), it does not matter, prefer (B), and strongly prefer (B). For each pair, programmers were allowed to provide an open-ended comment for why they chose the particular code pattern.

\textbf{Presented edit patterns} We presented programmers with the following nine edit patterns:


\begin{enumerate}[label=\fcolorbox{black}{white}{\normalsize EP\protect\twodigits{\arabic*} }, leftmargin=1.5cm,itemsep=0.1pt,labelsep=0.1cm]


\item (Performance) \textbf{Use characters instead of single-character strings.} In Java, we can represent a character both as a \code{String} or a character. For operations such as appending a value to a \code{StringBuffer}, representing the value as a character improves performance---e.g.,
change \texttt{sb.append("a")} to \texttt{sb.append('a')}. This edit improved the performance of some operations in the Guava project by 10-25\%~\cite{CMGUAVA12}.

\item (Error Prone) \textbf{Prefer string literal in \code{equals} method.} Invoking \code{equals} on a \code{null} variable 
causes a \code{NullPointerException}. 
When comparing the value in a string variable to a string literal, programmers can overcome this exception by invoking the \code{equals} method on the string literal since  the \code{String} \code{equals} method checks whether the parameter is \code{null}---for example, using \texttt{"str".equals(s)} instead of \texttt{s.equals("str")}.

\item (Performance) \textbf{Avoid \code{FileInputStream} and \code{FileOutputStream}}. These classes override the \code{finalize} method. As a result, their objects are only cleaned when the garbage collector performs a sweep~\cite{URLDZONEINPUTSTREAM}. Since Java 7, programmers can use \code{Files.newInputStream} and \code{Files.newOutputStream} instead of \texttt{FileInputStream} and \texttt{FileOutputStream} to improve performance as recommended in this Java JDK bug-report~\cite{URLJAVAJDK}.


\item (Best practices) \textbf{Use the collection \code{isEmpty} method rather than checking the size.} Using the method \code{isEmpty} to check whether a collection is empty is preferred to checking that the size of the collection is zero. For most collections these constructions are equivalent, but for others computing the size could be expensive. For instance, in the class \code{ConcurrentSkipListSet}, the size method is not constant-time~\cite{URLORACLECONCURRENTSKIPLISTSET}. Thus, prefer \texttt{list.isEmpty()} to \texttt{list.size() == 0}. This edit pattern is included in the PMD catalog of rules.

\item (Multithreading) \textbf{Prefer \code{StringBuffer} to \code{StringBuilder}.} These classes have the same API, but \code{StringBuilder} is not synchronized. Since synchronization is rarely used~\cite{URLORACLESTRINGBUILDER}, \code{StringBuilder} offers high performance and is designed to replace \code{StringBuffer} in single threaded contexts~\cite{URLORACLESTRINGBUILDER}.

\item (Code Style) \textbf{Infer type in generic instance creation.} Since Java 7, programmers can replace type parameters to invoke the constructor of a generic class with an empty set (\code{<>}), diamond operator~\cite{URLORACLETYPEINFE} and allow inference of type parameters by the context. This edit ensures the use of generic instead of the deprecated raw types~\cite{URLRAWTYPES}. 
The benefit of the diamond operator is clarity since it is more concise.

\item (Design) \textbf{Remove raw types.}  Raw types are generic types without type parameters and were used in versions of Java prior to 5.0. They ensure compatibility with pre-generics code. Since type parameters of raw types are unchecked,  unsafe code is caught at runtime~\cite{URLRAWTYPES} and the Java compiler issues warnings for them~\cite{URLRAWTYPES}. Thus, prefer \texttt{List<String> a = new ArrayList<>()} to \texttt{List<String> a = new ArrayList()}.

\item (Error Prone) \textbf{Field, parameter, and variable could be \code{final}.} The \code{final} modifier can be used in fields, parameters, and local variables to indicate they cannot be re-assigned~\cite{URLJAVAPRACTICESFINAL}. This edit improves clarity and it helps with debugging since
it shows what values will change at runtime. In addition, it allows the compiler and virtual machine to optimize the code~\cite{URLJAVAPRACTICESFINAL}. The edit pattern that adds the \code{final} modifier is included in PMD catalog of rules~\cite{pmd}. IDEs such as Eclipse~\cite{eclipse} and NetBeans~\cite{netbeans} can be configured to perform this edit automatically on saving.

\item (Error Prone) \textbf{Avoid using strings to represent paths.} Programmers sometimes use \code{String} to represent a file system path even though some classes are specifically designed for this
task---e.g., \texttt{java.nio.Path}. In these cases, it is useful to change the type of the variable to \code{Path}. First, strings can be combined in an undisciplined way, which can lead to invalid paths.  Second, different operating systems use different file separators, which can cause bugs. 
Since detecting this pattern requires a non-trivial analysis, code analyzers do not include it as a rule. Thus, use \code{Path path} over \code{String path}.

\end{enumerate}

\textbf{Analysis} We treated the Likert-type responses as ordinal data and applied a one-sample Wilcoxon signed-rank test to identify statistical differences for each of the nine edit patterns ($\alpha = 0.05$). Specifically, the null hypothesis is that the responses are not statistically different and symmetric around the default value (``it does not matter''). Rejecting the null hypothesis implies that programmers have a non-default preference for one code pattern. Because multiple comparisons can inflate the false discovery rate, we compute adjusted p-values using a Benjamini-Hochberg correction~\cite{10.2307/2346101}. To ease interpretation, we present the results for each pair as a net stacked distribution.

\subsection{Pull Request Validation}
\label{s:pullrequest}

To further assess the perceived usefulness of our edit patterns, and validate their acceptance, we submitted pull requests to GitHub projects containing the nine quick fixes from our survey.

\textbf{Project selection} From the nine GitHub projects (\cref{t:projects}), we selected five projects that actively considered pull requests (Ant, Error Prone, ExoPlayer, Giraph, and Gson). We supplement these projects those of four popular code analyzer tools
(Checkstyle, PMD, SonarQube, and Spotbugs) with the expectation that reviewers of these pull requests could capably assess the usefulness of the proposed quick fixes.  Thus, we selected a total of nine projects to submit pull requests.

\textbf{Pull requests} 
 We deliberately submitted pull requests that applied locally to a single region of code within a single file to minimize confounds that would be otherwise introduced in large pull requests. In total, we submitted~\pullrequests{} pull requests across all projects.

\textbf{Analysis} We recorded the status of the pull requests as either open (not yet accepted into the project code), merged (accepted into the project code), or rejected (declined to accept into the project code). We describe these pull-request submissions through basic descriptive statistics.

\section{Results}
\label{sec:results}

\subsection{\RQA (RQ1)}
\label{s:editpatterns}

\cref{t:rulecategories} characterizes the identified edit patterns and labels them according to the PMD rule-sets. The discovered patterns covered seven of the eight PMD categories, and only ``Security'' was not represented. The most common rule-sets---with roughly equal frequencies---were ``Design'' (22), ``Best Practices'' (19), ``Performance'' (13), and ``Error Prone'' (12). The results suggest that \technique{} is effective at discovering quick fixes across a spectrum of rule-sets.

Cohen's $\kappa$ found ``very good''~\cite{10.2307/2529310} agreement between the raters for these rule-sets ($n = 89$, $\kappa = 0.82$), with disagreement being primarily attributable to whether an edit pattern is ``Best Practice'' or ``Error Prone.''

Finally,~\newpatterns{}/\edits{} patterns were classified as new ones---i.e.,
they were not implemented as quick fixes in existing tools.

\begin{result}
\technique{} could automatically discover \edits{} edit patterns that covered 7/8 PMD categories. 
\percentneepatterns{} of the discovered patterns did not appear in existing tools.
\end{result}

\subsection{\RQB (RQ2)}
\label{sec:result:rq2}




\newcommand{\likert}[4]{
\begin{minipage}[l]{\textwidth}
  \begin{tikzpicture}[xscale=0.01, yscale=0.3]

    \node at (-100,0) {};
    \node at (100,0) {};

    \filldraw[color=Red] (#1, 0.0) rectangle (#2, 1.0);
    \filldraw[color=Pink] (#2, 0.0) rectangle (0, 1.0);
    \filldraw[color=LightGreen] (0, 0.0) rectangle (#3, 1.0);
    \filldraw[color=Green] (#3, 0.0) rectangle (#4, 1.0);

    \draw (0,0) -- (0, 1);

  \end{tikzpicture}
\end{minipage}
}

\begin{table}[!t]
\centering
\caption{Programmer preferences for edit patterns\label{tab:surveypref}}
\begin{threeparttable}

\begin{tabularx}{\linewidth}{lrrrrX}
\toprule
& & \multicolumn{3}{c}{\textbf{Likert Resp. Pct\tnote{1}}}\\
\cmidrule{3-5}
\textbf{Pattern} & 
\textbf{Adj-$p$\tnote{2}} &
\textbf{B} &
\textbf{N} &
\textbf{QF} &
\textbf{Distribution\tnote{3}}\\
    
\midrule

& & & & &

\begin{minipage}[l]{\textwidth}
  \begin{tikzpicture}[xscale=0.01, yscale=0.3]

    \node at (-100,0) {};
    \node at (100,0) {};

    \draw (-100,0) -- (100,0);
    \draw (-100,-0.25) -- (-100, 0.25);
    \draw (100,-0.25) -- (100, 0.25);
    \draw (-50,-0.25) -- (-50, 0.25) node[above] {\footnotesize 50\%};
    \draw (50,-0.25) -- (50, 0.25) node[above] {\footnotesize 50\%};
    \draw (0,-0.25) -- (0, 0.25) node[above] {\footnotesize 0\%};
  \end{tikzpicture}
\end{minipage}\\

EP1 & .01 & 49\% & 18\% & 33\%&
\likert{-48.78}{-34.15}{26.83}{32.93}\\

EP2 & $< .001$ & 33\% & 2\% & 65\%&
\likert{-33.54}{-14.63}{22.56}{64.63}\\

EP3 & .03 & 36\% & 18\% & 46\%&
\likert{-36.59}{-31.10}{29.88}{45.73}\\

EP4 & $< .001$ & 2\% & 5\% & 93\%&
\likert{-1.83}{-1.83}{40.24}{92.68}\\

EP5 & $< .001$ & 7\% & 21\% & 72\%&
\likert{-7.32}{-6.10}{28.05}{71.95}\\

EP6 & $< .001$ & 10\% & 1\% & 89\%&
\likert{-9.76}{-7.32}{20.73}{89.02}\\

EP7 & $< .001$ & 19\% & 6\% & 75\%&
\likert{-18.96}{-14.63}{25.61}{75.00}\\

EP8 & $< .001$ & 33\% & 12\% & 55\%&
\likert{-33.54}{-25.00}{33.54}{54.88}\\

EP9 & $< .001$ & 15\% & 16\% & 69\%&
\likert{-15.24}{-14.02}{46.95}{68.90}\\
\bottomrule
\end{tabularx}

\begin{tablenotes}
\item[1] Likert-type item responses: Strongly prefer or prefer baseline (B), Neutral (N), Strongly prefer or prefer quick fix (QF).
\item[2] Adjusted $p$-value after Benjamini-Hochberg correction.
\item[3] Net stacked distribution removes the Neutral option and shows the skew between baseline and quick fix preferences. \tikz \filldraw[color=Red] (0,0) rectangle (5pt,5pt); Strongly prefer baseline, \tikz \filldraw[color=Pink] (0,0) rectangle (5pt,5pt); Prefer baseline, \tikz \filldraw[color=LightGreen] (0,0) rectangle (5pt,5pt); Prefer quick fix, \tikz \filldraw[color=Green] (0,0) rectangle (5pt,5pt); Strongly prefer quick fix.
\end{tablenotes}

\end{threeparttable}

\end{table} 

A summary of the survey results is presented in \cref{tab:surveypref}. The Wilcoxon signed-rank test identified a significant difference in preference---after Benjamini-Hochberg adjustment---for all nine edit patterns (at $\alpha = 0.05$). With the exception of EP1, ``String to character,'' programmers preferred the quick fix version of the code from \technique{}---the strength of this preference is visible through the presentation of the net stacked distribution.

To understand why programmers rejected EP1, we examined the optional programmer feedback for this edit pattern. We found that although programmers recognized that passing a character would have better performance, ``slightly more efficient,'' and requiring ``less overhead,'' these benefits were not significant enough to outweigh readability or consistency. For example, five programmers reported that since the name of the class is \code{StringBuffer}, it's more consistent to always pass in a \code{String}, even if a character would be more efficient. Other programmers reported that always passing in a \code{String} is just ``easier mentally'' and requires ``less cognitive load.''


\begin{result}
Programmers preferred the quick fixes suggested by \technique{} for eight of the nine edit patterns.
\end{result}

\subsection{\RQC (RQ3)}
\label{sec:result:rq3}

\newcommand{\praccept}[1]{\tikz \filldraw[color=Green] (0,0) rectangle (5pt,5pt); #1}

\newcommand{\prreject}[1]{\tikz \filldraw[color=Red] (0,0) rectangle (5pt,5pt); #1}

\newcommand{\propen}[1]{\tikz \filldraw[color=Grey] (0,0) rectangle (5pt,5pt); #1}

\begin{table}[!t]
\centering
\caption{Pull request submissions to projects on GitHub\label{t:pullrequests}}
\begin{threeparttable}
\begin{tabularx}{\linewidth}{p{1cm}rrX}
\toprule
\textbf{Pattern} & \textbf{Accept} & \textbf{(\%)} & \textbf{Status}\tnote{1}\\
\midrule
EP1 & 1 & (33\%) & \praccept{PMD} \propen{Ant} \prreject{SonarQube}\\\addlinespace

EP2 & 2 & (67\%) & \praccept{Ant} \praccept{ExoPlayer}\newline\prreject{Error Prone} \\\addlinespace

EP3 & --- & --- & \propen{Giraph}\\\addlinespace

EP4 & 2 & (100\%) & \praccept{Ant} \praccept{PMD}\\\addlinespace

EP5 & 1 & (33\%) & \praccept{CheckStyle} \propen{Ant}\newline\prreject{Spotbugs}\\\addlinespace

EP6 & --- & --- & \propen{Giraph}\\\addlinespace

EP7 & --- & --- & \propen{Gson}\\\addlinespace

EP8 & 0 & (0\%) & \prreject{Gson}\\\addlinespace

EP9 & --- & --- & \propen{Giraph}\\\addlinespace
\midrule
\textbf{Total}\tnote{2} & 6 / 10 & (60\%)\\
\bottomrule
\end{tabularx}
\begin{tablenotes}
\item [1] \praccept{Accepted pull request}, \prreject{Rejected pull request}, \propen{Open pull request}.
\item [2] Acceptance rate is calculated as accepted pull requests against accepted and rejected pull requests. Open pull requests are not included in this calculation.
\end{tablenotes}
\end{threeparttable}
\end{table}

Of the~\pullrequests{} pull requests we submitted to GitHub projects,~\acceptpullrequeststext{}
of these were accepted,~\rejectedpullrequeststext{} were rejected, and the remaining are open as of the time of this writing (\cref{t:pullrequests}).\footnote{Links to GitHub pull requests temporarily removed for blind review.}

SonarQube rejected a pull request for EP1, suggesting that changing a \code{String} to a \code{Character} is purely pedantic. However, they welcomed additional evidence of the performance benefits and would be willing to reconsider given such evidence. Error Prone programmers indicated that EP2---using the \code{equals} method of a string literal---was generally useful but not for the particular use case to which we submitted the pull request. SpotBugs rejected the pull request for EP5---using \code{StringBuilder} instead of \code{StringBuffer}---because a maintainer did not want to unnecessarily make the commit history noisy unless the change was in a performance critical path. Finally, Gson rejected a EP8 pull request for adding \code{final} to a parameter: namely, because their IntelliJ already highlights locals and parameters differently depending on whether they are assigned to. In other words, the programmers already use an alternative means to communicate information about effectively \code{final} parameters.


\begin{result}
Projects accepted~\percentacceptedPullRequests{} of the pull requests for the \technique{} quick fixes.
\end{result}

\section{Limitations}
\label{sec:limitations}

Each of the three studies have limitations, which we describe in this section.

\textbf{Evaluation of \technique{}} We considered only single-location edit patterns, which are representative of most \quickpl{} in code analyzers. However, a current limitation of the technique is that it cannot identify dependent patterns---for example, when both return type of the method and the return statement must change together. Another limitation of our approach is that we only evaluated Java-based GitHub projects; both the choice of language and the choice of projects influence the quick fixes we identified. A threat to construct validity is that it is difficult to exhaustively determine if a discovered quick fix is actually novel. To mitigate this threat, we catalogued popular code analysis tools and conducted searches to find quick fixes. Similarly, given that the categorization of quick fixes involves human judgments, our results (\cref{t:rulecategories}) should be interpreted as useful estimators for \technique{}.

\textbf{Survey study.} The survey employed purposive---that is, non-random---sampling and evaluated only a limited number of quick fixes that do not exhaustively cover the entire design space of quick fixes. Thus, a threat to external validity is that we should be careful and avoid generalizing the results from this survey to all quick fixes. Moreover, participants in the survey self-reported their experience and may not necessarily have been experts. A construct threat within this study is that programmers are not directly evaluating quick fixes: rather, they are being asked to evaluate two different code snippets---essentially, the input and output to an editor pattern. Responses and explanations for their preferences may have been different had they been explicitly told to evaluate the quick fixes directly.

\textbf{Pull request validation.} The choice of projects we submitted pull requests to also influences the acceptance or rejection of the pull requests. As discussed in \cref{sec:discussion}, a construct validity threat is that pull requests are not the typical environment through which programmers apply quick fixes. Consequently, the acceptance and rejection of pull requests are not representative of how programmers would actually apply quick fixes within their development environment. Despite this limitation, the study validates that discovered quick fixes are adopted by projects, and provides explanation in cases for when they are not.

\section{Discussion}
\label{sec:discussion}

\textbf{Generating executable rules (RQ1)}
\label{s:transformation}
\technique{} generates AST patterns, but ideally one wants
to  generate executable \quickpl{} that can be 
added to code analyzers.
When possible, \technique{} compiles the generated patterns
to executable 
Refaster rules.
Refaster~\cite{WA13SCAL} is a rule-language used in the
code analyzer Error Prone~\cite{ErrorProne}.
A Refaster edit pattern is described
using 
\rone a before template to pattern-match target locations,
and \rtwo an after template to specify how these locations are transformed, which are similar to the before and after
templates $\tau_i$ and $\tau_o$ used by our rules.
In general, our rules cannot be always expressed as Refaster ones. 
In particular, Refaster cannot describe edit patterns that require AST node types. 
For instance,
the edit pattern in Figure~\ref{fig:template} requires knowing that an AST node is a \code{StringLiteral} and this AST type cannot be 
inspected in Refaster.
In addition, Refaster can only modify expressions that appear inside a method body---e.g., 
Refaster cannot modify global field declarations.
In the future, we plan to implement an extension of Refaster that can execute all rules in our format.

\textbf{Programmers consider trade-offs when applying quick fixes (RQ2)} The feedback from programmers within our survey study suggests trade-offs that programmers consider when making judgments about applying quick fixes. For example, we saw in \cref{sec:result:rq2} that for EP1 some programmers preferred the version of the code with worse performance  primarily because they valued \emph{consistency} and \emph{reduction in cognitive load} over what they felt was relatively small performance improvements.

But even when programmers significantly preferred the quick fixes from \technique{}, they carefully evaluated the trade-offs for their decision. For example, consider EP2, in which the quick fix suggests using the \code{equals method} on string literals to prevent a \code{NullPointerException}. Programmers recognized this benefit but also argued that the baseline version had better \emph{readability}. As one programmer notes, when given a variable, they felt it more natural to say, ``if variable equals value'' than ``if value equals variable.''

Programmers also indicated \emph{unfamiliarity} with new language features as a reason to avoid the quick fix version of the code, for example, in EP3---which uses the newer API of \code{Files.newInputStream} rather than \code{FileInputuStream}. One programmer noted that newer APIs embed ``experience about the shortcomings of the old API'' but they were also hesitant to use this version of the code without understanding what the shortcomings actually were.

Finally, using the diamond operator (\code{<>}) in EP6 makes the code simpler, concise, and more readable. Nevertheless, 19\% of participants still preferred or strongly preferred the less concise baseline version of the code. One programmer suggested \emph{compatibility} with old versions as a reason for this decision: although the diamond operator has several benefits, it cannot be supported if there is a need to target older versions of the Java specification.

Thus, even when automated techniques such as \technique{} discover useful quick fixes, the feedback from our survey suggests that it is also important to provide programmers with rationale for why and when the quick fix should be applied. A first-step towards providing an initial rationale can be to situate quick fixes within an existing taxonomy, as we did with our discovered quick fixes in \cref{t:rulecategories}.

\textbf{Barriers to accepting pull requests (RQ3)} Although our survey indicated that programmers significantly preferred the quick fixes identified by \technique{}, maintainers of projects in GitHub did not always accept the corresponding pull requests.

In our pull request study in \cref{sec:result:rq3}, the comments from project maintainers suggested reasons for declining a quick fix, even when they recognized that the fixes would be \emph{generally} useful. For instance, the maintainers of SonarQube did not want to incorporate the quick fixes because it would make the commit history more noisy. Spotbugs was concerned about adopting quick fixes without sufficient testing because the fixes might introduce \emph{regressions}, or behave unexpectedly in different JVM implementations. 
Other projects like Error Prone have adopted  conventions across their entire code base. Unless these quick fixes are applied universally across the project, such inertia makes it unlikely that these projects would adopt a one-off fix---for example, EP2.

Our analysis suggests that \emph{when} and \emph{how} a quick fix is surfaced to the developer is important to its acceptance. It is possible these maintainers would have applied these rejected quick fixes had they been revealed \emph{as} they were writing code, rather than after the fact.

\section{Related Work}

\textbf{Systems for mining edit patterns from code}
Molderez et al.~\cite{MO17MINI} learn AST-level tree transformations as sets of tree edits
and
used them to automate repetitive code edits.
Since the same pattern can 
be described with different sets of tree edits, two patterns that are deemed 
equivalent by \technique{} can be deemed nonequivalent in~\cite{MO17MINI}. 
Finally, the edits learned by it are 
not publicly available and were not evaluated through a survey.
Brown et al.~\cite{Brown17} learn token-level syntactic transformations---e.g., delete a variable or insert
an integer literal---from online code commits
to generate mutations for
mutation testing. Unlike \technique{}, they can only mine token level transformations
over a predefined set of syntactic constructs and cannot unify across multiple
concrete edits.
Negara et al.~\cite{NE14MINI} mine code interactions directly from the IDE to detect 
repetitive edit patterns and find 10 new
refactoring patterns.
\technique{} does not use continuous interaction data from an IDE and only makes use of public data available in online repositories. 
Other tools~\cite{LI18MINI,SO18DISS} mine fine-grained repair templates from StackOverflow 
and the Defect4j bug data-set and therefore differ from
\technique{}.
In summary, our paper differs from prior work in that 
\rone \technique{}  mines new edit patterns in a sound and complete fashion using a rich syntax of edit patterns, 
\rtwo we assessed the quality of the learned patterns and the corresponding \quickpl{} through a formal evaluation.




\textbf{Learning transformations from examples}
Several techniques use user-given examples 
to learn repetitive code edits for refactoring~\cite{RO17LEARN,ME13LASE,AN08GENE}, 
for removing code clones~\cite{ME15RASE}, for removing defects from code~\cite{KE11DESI}, for learning ways to fix command-line errors~\cite{DAntoniSV17}, and for performing code
completion~\cite{GA10ASTU,HI12ONTH}.
All these techniques rely on user-given examples
that
describe the same intended transformation
or on curated labeled data. This extra information allows the tools to perform more informed
types of rule extraction. Instead, \technique{} uses fully unsupervised learning and receives
concrete edits as input that may or may not describe useful transformations.


\textbf{Program repair}
Some program repair tools  learn useful fixing strategies by mining 
curated sets of bug fixes~\cite{NG10RECU}, user interactions with a debugger~\cite{JE09BUGF},
human-written patches~\cite{KI13AUTO,Long:2016:APG:2914770.2837617},
and bug reports~\cite{LI13R2FI}.
All these tools either rely on a predefined set of patches or learn patches from supervised data---e.g.,
learn how to fix a \code{NullPointerException} by mining all concrete edits that were performed to fix
 that type of exception.
Unlike these techniques,  \technique{} analyzes  unsupervised sets of concrete code edits and uses a sound and complete technique 
for mining a well-defined family of edit patterns. Moreover, \technique{} learns arbitrary \quickpl{} 
that can improve code quality
not just ones used to repair buggy code.
Since we do not have a notion of correct edit pattern---i.e., there is no bug to fix---we also
analyze the usefulness of the learned \quickpl{} through a comprehensive evaluation
and user study, a component
that is not necessary for the code transformations used in program repair.

\section{Conclusion}

We presented \technique{}, a technique for automatically 
discovering common Java code edit patterns in online code repositories.
\technique{} 
\rone
identifies edits by comparing consecutive 
revisions in  code repositories, 
\rtwo clusters edits into sets that 
can be abstracted into the same edit pattern, and
we used  \technique{} to mine \quickpl{} from \projectsspell{}  popular GitHub Java
projects and \technique{} successfully learned
\edits{} edit patterns that appeared in more than 
three projects. 
To assess whether programmers would like to apply these \quickpl{} to their code, we
performed an online survey with \answers{} programmers showing 9 of our \quickpl{}. 
Overall, programmers supported \psupported{}
of our \quickpl{}. 
We also issued pull requests in various repositories 
and \ppullrequests{} were accepted so far.

The results of this work have several implications for toolsmiths. First, \technique{} can be used to efficiently collect patterns and their usages in actual repositories and enable toolsmiths to make informed decisions about which quick fixes to prioritize based on patterns programmers actually apply in practice. Second, \technique{} allows toolsmiths to discover new quick fixes, without needing their users to explicitly submit quick fix suggestions. Third and finally, the results of this work suggest several logical and useful extensions to further aid toolsmiths---e.g., supporting more complex patterns that appear in code analyzers but are currently beyond the capabilities of \technique{} and designing techniques for automatically extracting executable quick fixes from mined patterns.

\bibliographystyle{IEEEtran}
\bibliography{ref}

\end{document}